\documentclass[pra,twocolumn, preprintnumbers,amsmath,amssymb, nofootinbib]{revtex4-1}

\usepackage{graphicx, color, graphpap}
\usepackage{amsmath}
\usepackage{enumitem}
\usepackage{amssymb}
\usepackage{amsthm}
\usepackage{pstricks}
\usepackage{float}
\usepackage{multirow}
\usepackage[hidelinks]{hyperref}
\usepackage{overpic}
\usepackage[T1]{fontenc}
\usepackage{algorithm}
\usepackage{algpseudocode}


\long\def\ca#1\cb{} 

\newcommand{\balpha}{\boldsymbol{\alpha}}
\newcommand{\bbeta}{\boldsymbol{\beta}}
\newcommand{\bgamma}{\boldsymbol{\gamma}}
\newcommand{\bpsi}{\boldsymbol{\Psi}}
\newcommand{\valpha}{\vec{\alpha}}
\newcommand{\perm}{P}

\newcommand{\dya}[1]{\ket{#1}\!\bra{#1}}
\newcommand{\dyad}[2]{\ket{#1}\!\bra{#2}}        

\newcommand{\Real}{{\rm Re}}
\newcommand{\Imag}{{\rm Im}}
\newcommand{\Tr}{{\rm Tr}}

\renewcommand{\geq}{\geqslant}

\newcommand{\mte}[2]{\langle#1|#2|#1\rangle }

\newcommand{\order}{n}
\newcommand{\qubits}{k}

\newtheoremstyle{example}{\topsep}{\topsep}%
{}
{}
{\bfseries}
{:}
{   }
{\thmname{#1}\thmnumber{ #2}}
\theoremstyle{example}

\theoremstyle{definition}

\input{Qcircuit}
\begin{document}

\title{Entanglement spectroscopy with a depth-two quantum circuit}

\author{Yi\u{g}it Suba\c{s}\i} 
\affiliation{Theoretical Division, Los Alamos National Laboratory, Los Alamos, NM 87545, USA.}

\author{Lukasz Cincio} 
\affiliation{Theoretical Division, Los Alamos National Laboratory, Los Alamos, NM 87545, USA.}

\author{Patrick J. Coles} 
\affiliation{Theoretical Division, Los Alamos National Laboratory, Los Alamos, NM 87545, USA.}

\begin{abstract}
Noisy intermediate-scale quantum (NISQ) computers have gate errors and decoherence, limiting the depth of circuits that can be implemented on them. A strategy for NISQ algorithms is to reduce the circuit depth at the expense of increasing the qubit count. Here, we exploit this trade-off for an application called entanglement spectroscopy, where one computes the entanglement of a state $| \psi \rangle$ on systems $AB$ by evaluating the R\'enyi entropy of the reduced state $\rho_A = {\rm Tr}_B(|  \psi \rangle \langle \psi |)$. For a $k$-qubit state $\rho(k)$, the R\'enyi entropy of order $n$ is computed via ${\rm Tr}(\rho(k)^{n})$, with the complexity growing exponentially in $k$ for classical computers. Johri, Steiger, and Troyer [PRB {\bf 96}, 195136 (2017)] introduced a quantum algorithm that requires $n$ copies of $| \psi \rangle$ and whose depth scales linearly in $k*n$. Here, we present a quantum algorithm requiring twice the qubit resources ($2n$ copies of $| \psi \rangle$) but with a depth that is independent of both $k$ and $n$. Surprisingly this depth is only two gates. Our numerical simulations show that this short depth leads to an increased robustness to noise.
\end{abstract}

\break
\newpage
\newpage
\maketitle

\section{Introduction}\label{sctintro}

Quantum computers promise exponential speedups for various applications, such as simulation of quantum systems \cite{feynman1982simulating}. Near-term devices, referred to as noisy intermediate-scale quantum (NISQ) computers \cite{preskill2018quantum}, are not yet in the regime of realizing these speedups, although quantum supremacy \cite{preskill2012quantum, neill2017blueprint} for a specially designed academic problem may be coming soon. Nevertheless, the question of what NISQ computers may be useful for remains an interesting one \cite{preskill2018quantum}.

Decoherence and gate fidelity continue to be important issues for NISQ devices~\cite{ball2018the}.  Ultimately these issues limit the depth of algorithms that can be implemented on these computers and increase the computational error for short-depth algorithms. Furthermore, NISQ computers do not currently have enough qubits, sufficient coherence times, and gate fidelities to fully leverage the benefit of quantum error-correcting codes~\cite{fowler2012surface, you2013simulating}. This highlights the need for strategies to reduce the depth of quantum algorithms in order to avoid the accumulation of errors~\cite{temme2017error}.

One such strategy notes that there is often a trade-off between the circuit depth and the number of qubits involved in one's algorithm \cite{broadbent2009parallelizing}. Namely, increasing the number of qubits can lead to shorter depth. Recently, industry quantum computers seem to be increasing their qubit counts relatively rapidly, although these qubits are noisy \cite{ball2018the}. So this strategy may be fruitful in the near term (i.e., before error correction is possible). A second strategy notes that quantum algorithms can be hybridized (i.e., made into quantum-classical algorithms) whereby part of the computation is done on a classical computer \cite{mcclean2016theory, cincio2018learning}. This reduces the load for the (error-prone) quantum computer.

In this paper, we employ both of these strategies to dramatically reduce the circuit depth for a particular application called entanglement spectroscopy \cite{li2008entanglement,Amico2008,horodecki2009quantum}. Here one computes the entanglement of a pure bipartite quantum state $\ket{\psi}$ on systems $AB$ by measuring various R\'enyi entropies of the reduced state $\rho_A = \Tr_B(\dya{\psi})$. The entanglement between subsystems $A$ and $B$ in $\ket{\psi}$ is completely characterized by the eigenvalues of $\rho_A$. Li and Haldane noted that the largest eigenvalues of $\rho_A$ contain more universal signatures than the von Neumann entropy alone \cite{li2008entanglement}. They introduced the concept of entanglement spectrum, writing $\rho_A = \exp(-H_E)$ as the exponential of the ``entanglement Hamilonian'' so that the largest eigenvalues correspond to the lowest energies of $H_E$. As noted in \cite{song2012bipartite}, the integer R\'enyi entropies of $\rho_A$ can be used to reconstruct the largest eigenvalues of $\rho_A$. 

Entanglement spectroscopy will be important in the future when quantum computers are large enough to perform quantum simulation of many-body systems \cite{islam2015measuring,linke2017measuring}. Imagine that $\ket{\psi}$ is the output of the simulation, and one wants to quantify the bipartite entanglement in this state. Since $\ket{\psi}$ is already in quantum form (as opposed to a vector of amplitudes, as one would store it on a classical computer), one can directly act with a quantum gate sequence and measurements on $\ket{\psi}$ to compute this figure-of-merit.   

\begin{figure}
\begin{center}
 \includegraphics[width=\columnwidth]{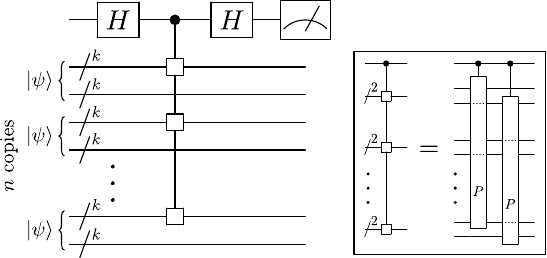}
\caption{Algorithm presented in \cite{johri2017entanglement} to compute $\Tr(\rho_A^{\order})$ for integer $n\geq 2$. Here, $\rho_A$ contains $\qubits$ qubits and is the reduced state of $\ket{\psi}$ (containing $2\qubits$ qubits). Two Hadamards sandwich a controlled-permutation gate acting on $n$ copies of $\ket{\psi}$. The controlled-permutation gate is expanded in the inset, for the special case of $\qubits =2$. Each controlled-$P$ gate is then decomposed into $n$ controlled-swaps, which in turn are written in terms of CNOTs and one-body gates \cite{shende2009cnot}. This shows that the algorithm's gate depth grows in proportion to $k*n$. Here we assumed for simplicity that the state $\rho_A$ contains half of the qubits in $\ket{\psi}$. The generalization to arbitrary bipartition is straightforward.}
\label{fig:troyer}
\end{center}
\end{figure}

The R\'enyi entropy of order $\order$ is defined as
\begin{align}\label{eqn1}
    S_{\order}(\rho) = \frac{1}{1-\order} \log \left(R_{\order}(\rho)\right)
\end{align}
where 
\begin{align}\label{eqn2}
    R_{\order}(\rho) = \Tr (\rho^{\order})\,,
\end{align}
and we consider $n\geq 2$ to be an integer in this work. Suppose that $\rho(\qubits)$ is a $\qubits$-qubit state. Since $\rho(\qubits)$ is a $2^{\qubits}\times 2^{\qubits}$ matrix, the complexity of computing $\rho(\qubits)^{\order}$ and hence $S_{\order}(\rho(\qubits))$ grows exponentially with $\qubits$ for a classical computer. In contrast, Johri et al.~\cite{johri2017entanglement} introduced a quantum algorithm that computes $S_{\order}(\rho(\qubits))$ with complexity growing bilinearly in $\qubits$ and $\order$, i.e., with the product $\qubits*\order$. Their algorithm (henceforth referred to as the JST algorithm) generalized the well-known Swap Test for computing purity $\Tr(\rho^2)$ and state overlap. That is, by replacing the controlled-swap operator in the Swap Test with a controlled-permutation operator, their algorithm can compute $\Tr(\rho^{\order})$ for integer $\order\geq 2$. This algorithm is shown in Fig.~\ref{fig:troyer}.

In this work, we propose an alternative quantum algorithm for entanglement spectroscopy. Our algorithm dramatically shortens the depth relative to that of Ref.~\cite{johri2017entanglement}, at the expense of requiring more qubits and more classical post-processing. Namely, the JST algorithm   requires $\order$ copies of $\ket{\psi}$, with a circuit depth growing with $\qubits*\order$. At the end a single ancilla qubit is measured to compute the expectation value of the Pauli-$Z$ operator. In contrast, our algorithm requires $2\order$ copies of $\ket{\psi}$, while our circuit depth is, surprisingly, independent of both $\qubits$ and $\order$. Furthermore, this depth is only two quantum gates. 
At the end all qubits are measured and the post-processing of our algorithm grows in proportion to $\qubits*\order$. In this way we have transferred some of the complexity from quantum into classical computation. For NISQ devices, it is always better to push complexity onto classical computers, which are essentially error free.

\begin{figure}
\begin{center}
 \includegraphics[width=\columnwidth]{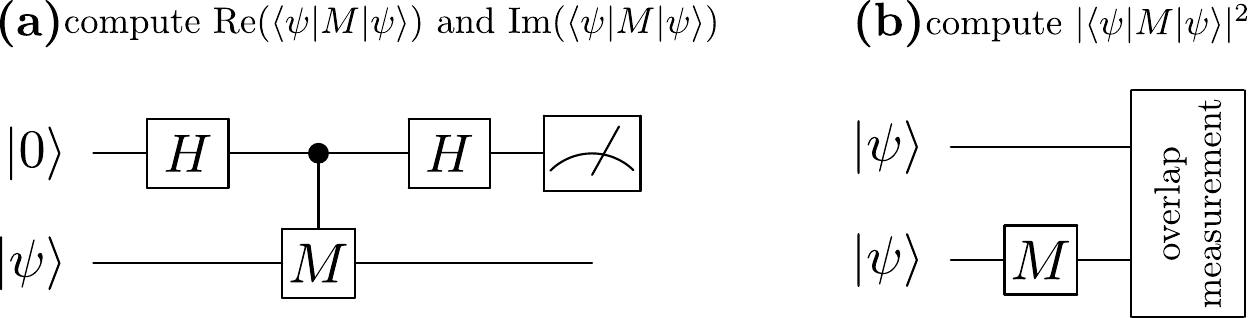}
\caption{Two different strategies for computing an operator's expectation value. (a) The Hadamard Test involves applying controlled-$M$ and requires one copy of $\ket{\psi}$ and one ancilla. By varying the final measurement in the $xy$ plane of the Bloch sphere, one can extract linear combinations of $\Real (\mte{\psi}{M})$ and $\Imag (\mte{\psi}{M})$. (b) Here we invoke a different algorithm that we call the Two-Copy Test, which requires two copies of $\ket{\psi}$. This algorithm applies $M$ to one of the two copies and then measures the overlap between the copies, giving $|\mte{\psi}{M} |^2$.}
\label{fig2}
\end{center}
\end{figure}

At the core of our algorithm is an alternative approach to computing the expectation value of an operator $M$. It is well known that the Hadamard Test can be used to find $\mte{\psi}{M}$ by implementing the controlled-$M$ gate, see Fig.~\ref{fig2}(a). In this work, we note that $|\mte{\psi}{M}|^2$ can be computed by implementing $M$ instead of controlled-$M$, if one allows for two copies of $\ket{\psi}$. We call the latter approach the Two-Copy Test, and it is depicted in Fig.~\ref{fig2}(b). For computing the R\'enyi entropies in Eq.~\eqref{eqn1}, $M$ is set to be the cyclic permutation operator acting on subsystem $A$ of the overall $AB$ system.

In what follows, we first give some background, including the connection between the permutation operator and the integer R\'enyi entropies as well as the connection between these entropies and the largest eigenvalues of the state. We then present our main result: a quantum circuit with a depth of two gates for computing the integer R\'enyi entropies. 
This is followed by a description of post-selection methods that might in some cases improve the accuracy of the results.
Next, we numerically simulate our circuit as well as the circuit in Fig.~\ref{fig:troyer}, and we discuss how our circuit leads to increased robustness to noise, particularly when the readout error is small compared to other sources of noise.
Finally, we compare hardware noise with statistical noise for our algorithm. 
Details about post-selection methods and numerical simulations are provided in Appendices.

\section{Background}\label{sctbackground1}

\subsection{R\'enyi entropies via the permutation operator}

Nonlinear functions of a state $\rho$ can be obtained by evaluating linear expectation values on multiple copies of $\rho$~\cite{brun2004measuring}. One of the most well-known examples is the swap trick~\cite{hastings2010measuring}, which uses two copies of $\rho$ to evaluate the purity:
\begin{align}\label{eqn3}
\Tr(\rho^{2}) = \Tr((\rho\otimes \rho) \text{SWAP})
\end{align}
where $\text{SWAP} = \sum_{j,k} \dyad{jk}{kj}$ is the swap operator. This trick generalizes to $\order \geq 2$ copies of $\rho$ as follows:
\begin{align}\label{eqn4}
\Tr(\rho^{\order}) = \Tr(\rho^{\otimes \order} P)
\end{align}
where
\begin{align}\label{eqn5}
\rho^{\otimes \order} = \rho \otimes \rho \otimes \ldots \otimes \rho\quad (\order \text{ times}).
\end{align}
Here
\begin{align}\label{eqn6}
P = \sum_{j_1,j_2, \ldots , j_{\order}} \dyad{j_{\order}j_1 j_2 \ldots j_{\order -1}}{j_1 j_2 \ldots j_{\order}}
\end{align}
is the cyclic permutation operator, permuting the $\order$ subsystems of $\rho^{\otimes \order}$.

An important property of $P$ is that it factorizes into a tensor product of permutation operators when acting on a tensor-product Hilbert space. To make this clear, let $P_{\qubits}^{(\order)}$ denote the permutation operator acting on $\order$ quantum systems, each of which is composed of $\qubits$ qubits. Suppose that $Q$ is a composite quantum system composed of $\order$ subsystems:
\begin{align}\label{eqn7a}
Q = Q^{(1)}Q^{(2)} \ldots Q^{(\order)}\,,
\end{align}
and that each $Q^{(i)}$ is composed of $\qubits$ qubits:
\begin{align}\label{eqn7}
Q^{(i)} = Q^{(i)}_{1}Q^{(i)}_{2} \ldots Q^{(i)}_{\qubits}
\end{align}
where $Q^{(i)}_{j}$ denotes the $j$-th qubit in the $i$-th subsystem, $Q^{(i)}$. Then, when $P_{\qubits}^{(\order)}$ acts on the Hilbert space associated with the $Q$ system, it can be written as
\begin{align}\label{eqn8}
P_{\qubits}^{(\order)} = P_{1}^{(\order)} \otimes P_{1}^{(\order)} \otimes \ldots \otimes P_{1}^{(\order)} \quad (\qubits \text{ times}),
\end{align}
provided that we order the qubits in the following way
\begin{align}\label{eqn9}
Q^{(1)}_1 Q^{(2)}_1 \ldots Q^{(\order)}_1 \ldots Q^{(1)}_{\qubits}Q^{(2)}_{\qubits} \ldots Q^{(\order)}_{\qubits}\,.
\end{align}
Note that $P_{1}^{(\order)}$ in \eqref{eqn8} is the operator that permutes $n$ subsystems each of which is composed of one qubit.

\subsection{Computing eigenvalues from R\'enyi entropies}

In order to exactly compute all eigenvalues $\{\lambda_i\}$ of the density matrix $\rho$ of a system of $k$ qubits, one needs to know all R\'enyi entropies up to order $2^k$. As noted in Refs.~\cite{song2012bipartite,johri2017entanglement} these quantities can be related to each other by the Newton-Girard Formula:
\begin{align}\label{eqn10}
    (x - \lambda_1)(x - \lambda_2) \ldots (x - \lambda_N) = \sum_{m=0}^N (-1)^{N-m} e_{N-m}x^m
\end{align}
where $N=2^k$ is the dimension of $\rho$ and
\begin{equation} \label{eqn11}
\begin{split}
e_0 &= 1\, ,\\
e_1 &= R_1\, ,\\
e_2 &= \frac{1}{2}(e_1 R_1 - R_2)\, ,\\
e_3 &= \frac{1}{3}(e_2 R_1 - e_1 R_2 + R_3)\, , \\
e_4 &= \frac{1}{4}(e_3R_1-e_2R_2+e_1R_3-R_4)\, , \\
    & \,\,\,\vdots \, .
\end{split}
\end{equation}
However, in most cases we are only interested in a small number of largest eigenvalues $\lambda_1\ge \lambda_2\ge \dots\ge \lambda_{n_{\text{max}}}$. 
In Refs.~\cite{song2012bipartite,johri2017entanglement} it was argued that an approximation to the $n_{\text{max}}$ largest eigenvalues of $\rho$ can be obtained by truncating the polynomial on the right-hand-side of Eq.~\eqref{eqn10} to that order and solving for the roots.
Using this method we can approximately compute $n_{\text{max}}$ largest eigenvalues of $\rho$ from the R\'enyi entropies of order up to $n_{\text{max}}$.

We remark that the eigenvalues obtained from Eq.~\eqref{eqn10} can be very sensitive to error in the coefficients $e_j$. To avoid this issue, Pichler et. al.~\cite{pichler2016measurement} proposed a measurement protocol in experiments with cold atoms to access the eigenvalues directly. However, their approach relies on the efficient implementation of many-qubit gates, which is possible with cold atoms. In this work we focus on hardware-agnostic algorithms for a quantum computer that is capable of implementing one- and two-qubit gates only.

\begin{figure}[t]
\begin{center}
 \includegraphics[width=\columnwidth]{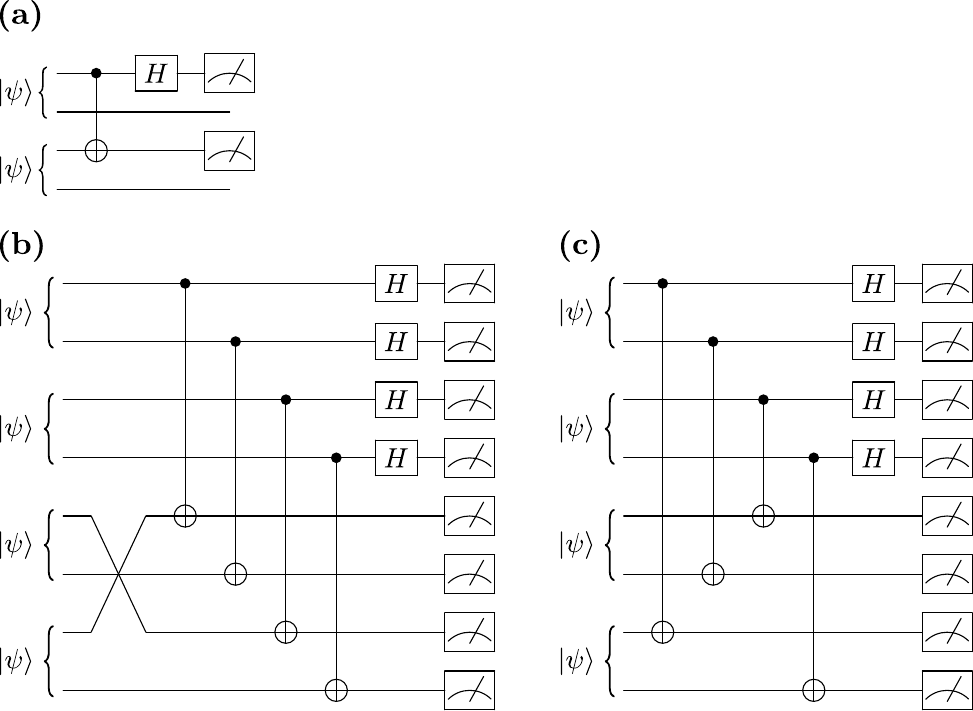}
\caption{Circuits for computing $\Tr(\rho_A^2)$ with a depth of two. The classical post-processing is not shown, but is discussed in the text. (a) Refs.~\cite{garcia2013swap,cincio2018learning} showed that the Bell-basis measurement on two copies of the state computes $\Tr(\rho_A^2$). (b) Our algorithm applied to $n=2$ and $k=1$, which is based on the Two-Copy Test shown in Fig.~\ref{fig2}. Namely, we feed in two copies of $\ket{\psi_2}:=\ket{\psi}\otimes \ket{\psi}$, i.e., four copies of $\ket{\psi}$, in order to compute $|\mte{\psi_2}{\text{SWAP}_A}|^2$, where $\text{SWAP}_A$ is the swap operator for the $A$ subsystems. This involves applying $\text{SWAP}_A$ to one copy of $\ket{\psi_2}$ and then measuring the overlap with the other copy of $\ket{\psi_2}$. The overlap measurement is the Bell-basis measurement, i.e., the same measurement employed in part (a) of this figure. (c) Note that the swap gate 
simply changes the targets of the subsequent CNOT gates in the circuit. Furthermore, note that all of the CNOTs and Hadamards can be performed in parallel, giving a circuit depth of two.}
\label{fig:tr_rho2}
\end{center}
\end{figure}

\section{Main Result}\label{sctmainresult}

Here we present our main result: a circuit for computing the integer R\'enyi entropies with a depth of only two quantum gates. We emphasize that our circuit does require access to the full pure state $\ket{\psi}$ in order to compute the R\'enyi entropies of the reduced state $\rho_A = \Tr_B(\dya{\psi})$. For readability, we first illustrate our circuit for the simplest case of computing purity for one-qubit states.

\subsection{Special case of $\order =2$, $\qubits =1$}

Suppose $\rho_A = \Tr_B(\dya{\psi})$ is a single-qubit state and one wishes to compute $\Tr(\rho_A^2)$. Previous work \cite{garcia2013swap,cincio2018learning} showed that this can be done via a Bell-basis measurement on two copies of the state, as depicted in Fig.~\ref{fig:tr_rho2}(a). This measurement involves applying a CNOT followed by a Hadamard on one of the copies. Finally one applies a classical post-processing as a simple dot product with the probability vector, i.e.,
\begin{align}\label{eqn12}
\Tr(\rho_A^2) = \vec{c}\cdot \vec{p}
\end{align}
where $\vec{p} = \{p_{00}, p_{10}, p_{01}, p_{11}\}$ is the probability vector for the measurement outcomes and $\vec{c} = \{1,1,1,-1\}$. 

For $n=2$ we recommend employing the aforementioned algorithm in Fig.~\ref{fig:tr_rho2}(a). Nevertheless, we show how the algorithm presented in this paper applies to the $\order=2$ case in Fig.~\ref{fig:tr_rho2}(b). The Two-Copy test in Fig.~\ref{fig2}(b) is the basis of our algorithm. We feed in two copies of $\ket{\psi_2}:=\ket{\psi}\otimes \ket{\psi}$, i.e., four copies of $\ket{\psi}$. We apply the swap operator $\text{SWAP}_A$ to one copy of $\ket{\psi_2}$, where the $A$ subscript indicates that the swap is being applied only to the $A$ subsystems. Then we measure the overlap with the other copy of $\ket{\psi_2}$, which gives:
\begin{align}\label{eqn13}
|\mte{\psi_2}{\text{SWAP}_A}|^2 = (\Tr(\rho_A^2))^2\,.
\end{align}
The proof of Eq.~\eqref{eqn13} is straightforward and is shown below in Eq.~\eqref{eqn16}. 

We emphasize that the implementation of the swap gate is trivial since its only effect is to change the ordering of the qubits, as shown in Fig.~\ref{fig:tr_rho2}(c). The same effect can be achieved by changing the indices of the target qubits in the CNOTs following the swap gate. 
The depth of the circuit in Fig.~\ref{fig:tr_rho2}(c) is due to the gates that compose the overlap measurement, and this depth is two gates, since the various CNOTs and Hadamards on distinct qubits can be parallelized.

The classical post-processing needed to obtain Eq.~\eqref{eqn13} from the measurement results in Fig.~\ref{fig:tr_rho2}(c) involves taking the dot product with the probability vector $\vec{p}$, as in Eq.~\eqref{eqn12}, but with 
\begin{align}\label{eqn14}
(\Tr(\rho_A^2))^2 = \vec{c}\cdot \vec{p},\quad
\vec{c} = \{1,1,1,-1\}^{\otimes 4}\,.
\end{align}
The form of $\vec{c}$ stated here requires one to reorder the qubits such that each qubit is grouped next to its overlap partner, i.e., each qubit controlling a CNOT in Fig.~\ref{fig:tr_rho2}(c) is immediately followed by the qubit being targetted by that CNOT. An explicit form of the post-processing in this case is thus given by:
\begin{equation}
    (\Tr(\rho_A^2))^2 = \sum_{j_1,\ldots,j_8} (-1)^{j_1j_7 + j_2j_6 + j_3j_5 + j_4j_8}
    p_{j_1,\ldots,j_8}\ .
\end{equation}

\begin{figure}[t]
\begin{center}
 \includegraphics[width=\columnwidth]{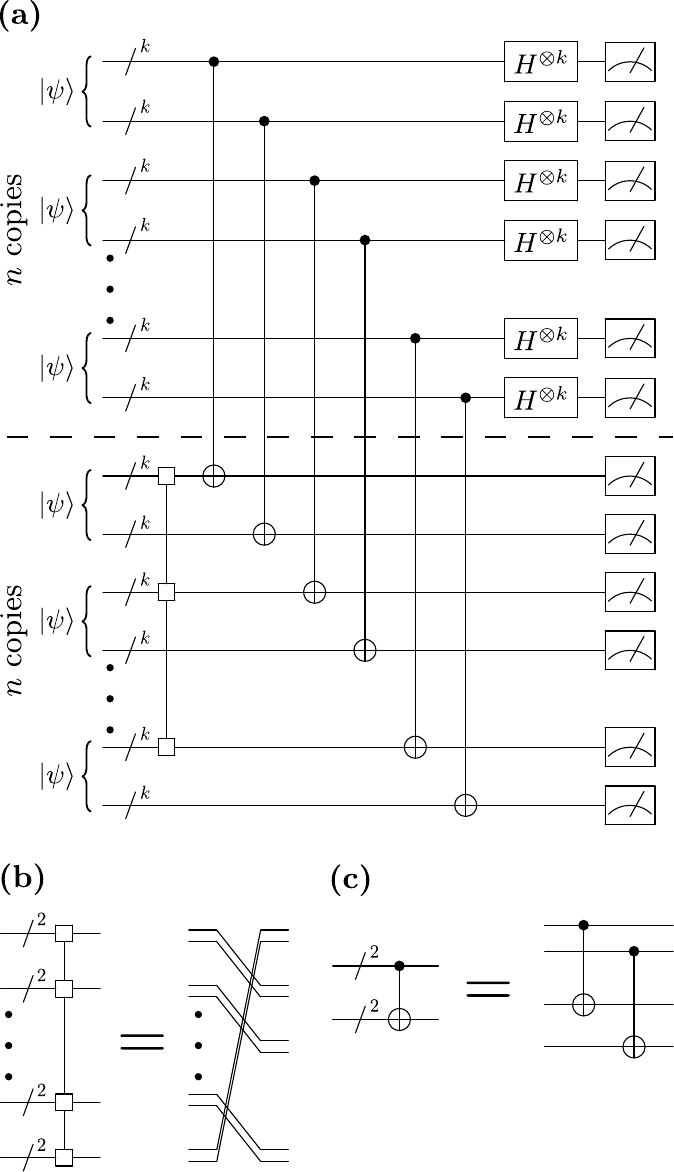}
\caption{Circuit for computing the integer R\'enyi entropies $S_n(\rho_A)$, or more precisely $\Tr(\rho_A^{\order})$, for $\order \geq 2$ where $\rho_A = \Tr_B(\dya{\psi})$.  The circuit acts on a total of $2\order$ copies of $\ket{\psi}$, or in other words, two copies of $\ket{\psi_{\order}}:= \ket{\psi}^{\otimes \order}$. We employ a compact notation for the cyclic permutation gate and for CNOT gates between multiple pairs of control and target qubits, respectively shown in (b) and (c) for the special case of $\qubits =2$.}
\label{fig:our_alg}
\end{center}
\end{figure}

\subsection{Circuit for general $\order$ and $\qubits$}

\begin{figure*}[t!]
\begin{center}
 \includegraphics[width=\linewidth]{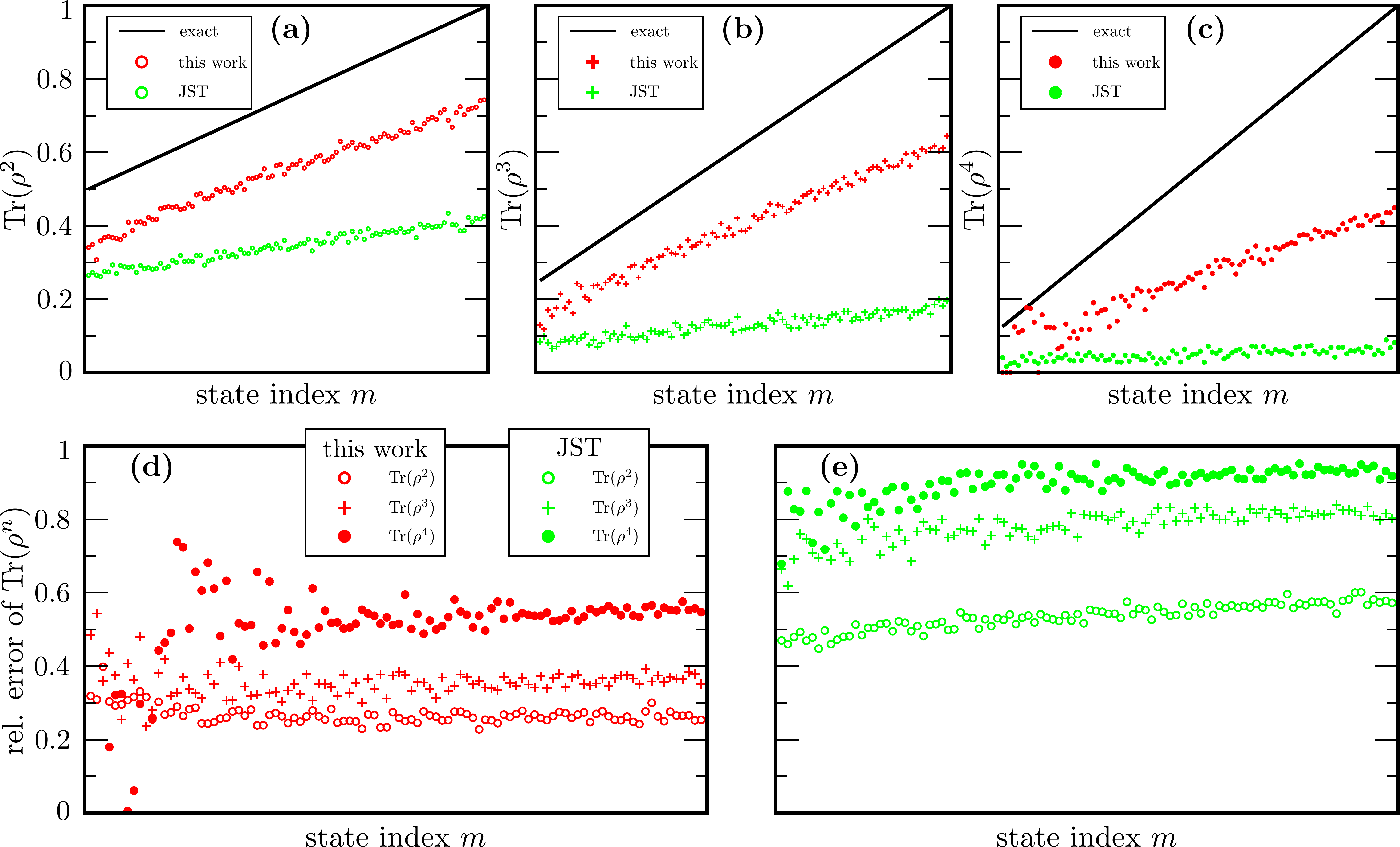}
\caption{Numerical simulation of our algorithm and the JST algorithm (Ref.~\cite{johri2017entanglement}) using IBM's QASM Simulator~\cite{qasm}. We have included relaxation, decoherence, readout and gate errors. All the noise parameters can be found in Appendix~\ref{sec:para}. The data points correspond to random states prepared according to Eq.~\eqref{eqn18}. Plots (a), (b), and (c) show the $n=2$, $n=3$, and $n=4$ cases respectively, including the exact curve (black), the data associated with the algorithm in Fig.~\ref{fig:our_alg} (red), and the data associated with the algorithm in Fig.~\ref{fig:troyer} (green). The relative error is plotted in (d) and (e) for the algorithms in Fig.~\ref{fig:our_alg} and Fig.~\ref{fig:troyer}, respectively.}
\label{fig:numerics}
\end{center}
\end{figure*}

Our main result is the circuit in Fig.~\ref{fig:our_alg}, which generalizes the special case shown in Fig.~\ref{fig:tr_rho2}(b). This gives a general algorithm for computing the integer R\'enyi entropies for $\order \geq 2$, for states with an arbitrary number of qubits. For simplicity, we assume the number of qubits in subsystems $A$ and $B$ are the same and equal to $k$, as the extension to arbitrary bipartitions is straightforward. Due to the circuit's generality, we introduce some compact circuit notation in Fig.~\ref{fig:our_alg}(b) and (c), respectively defining the permutation gate and CNOT gates between multiple pairs of control and target qubits.

The gate complexity (the total number of gates) of the algorithm is $4kn$. This is composed of $2kn$ CNOT gates and $2kn$ Hadamard gates.

Surprisingly, the circuit depth is independent of the problem size, i.e., independent of both $\order$ and $\qubits$. One can see this by noting that: (1) the cyclic permutation gate, shown in Fig.~\ref{fig:our_alg}(b), does not add to the circuit depth since it just reorders the qubits, and (2) the CNOT and Hadamard gates on distinct qubits can be parallelized. The result is a circuit with a depth of two quantum gates.

As noted earlier, the conceptual basis of our algorithm is the Two-Copy Test from Fig.~\ref{fig2}(b). We prepare two copies of the state $\ket{\psi_{\order}}:= \ket{\psi}^{\otimes \order}$. We apply the permutation gate $P_A$ to one of these copies, where the $A$ subscript indicates that the permutation is applied only to the $A$ subsystems. Then we measure the overlap with the other copy, giving
\begin{align}\label{eqn15}
|\mte{\psi_{\order}}{P_A}|^2 = (\Tr(\rho_A^{\order}))^2\,.
\end{align}
Taking the logarithm of Eq.~\eqref{eqn15} and dividing by $2(1-\order)$ then gives the R\'enyi entropy $S_{\order}(\rho_A)$. The proof of Eq.~\eqref{eqn15} is simply
\begin{align}
\nonumber
\mte{\psi_{\order}}{P_A} &= \Tr_{AB}(\dya{\psi_{\order}}P_A)\\
\nonumber
&= \Tr_{A}(\rho_A^{\otimes \order}P_A)\\
\label{eqn16}
&= \Tr (\rho_A^{ \order})\,.
\end{align}

The classical post-processing needed to obtain Eq.~\eqref{eqn15} from the measurement results is the generalization of what was discussed previously in Eq.~\eqref{eqn14}, namely
\begin{align}\label{eqn17}
(\Tr(\rho_A^n))^2 = \vec{c}\cdot \vec{p},\quad
\vec{c} = \{1,1,1,-1\}^{\otimes 2kn}\,.
\end{align}
Again, the form of $\vec{c}$ here requires a special ordering of the qubits, whereby each qubit that controls a CNOT in Fig.~\ref{fig:our_alg} is followed by the target qubit for that CNOT. Hence each vector $\{1,1,1,-1\}$ is associated with a pair of qubits, and there are a total of $2kn$ qubit pairs in Fig.~\ref{fig:our_alg}.

\section{Post-selection Methods}

Ref.~\cite{linke2017measuring} implemented the JST algorithm to compute $R_2$ for a one-qubit subsystem ($k=1$). There it was pointed out that if all qubits in Fig.~\ref{fig:troyer} are measured at the end of the computation (as opposed to measuring the ancilla qubit alone), some outcomes would be forbidden by the symmetries of the problem. If such outcomes are measured, it can be concluded that an error has occurred. By discarding such data points, they were able to improve their results, i.e., obtain a more accurate value for $R_2$. 

In Appendix~\ref{sec:johripost} we generalize the post-selection method employed in Ref.~\cite{linke2017measuring} for the JST algorithm. Our generalization works for all orders $n$ and number of subsystem qubits $k$, and in particular allows us to study the utility of post-selection for $n>2$ in the next section. We found the complexity of this post-selection method to be $O(k\, n \log n)$. 

In Appendix~\ref{sec:ourpost} we present a post-selection method for our algorithm that likewise can lead to more accurate results. It has complexity $O(k\,n^2)$. The effect of post-selection on the quality of results is analyzed numerically in the next section.  

\begin{figure*}[t]
\begin{center}
 \includegraphics[width=\linewidth]{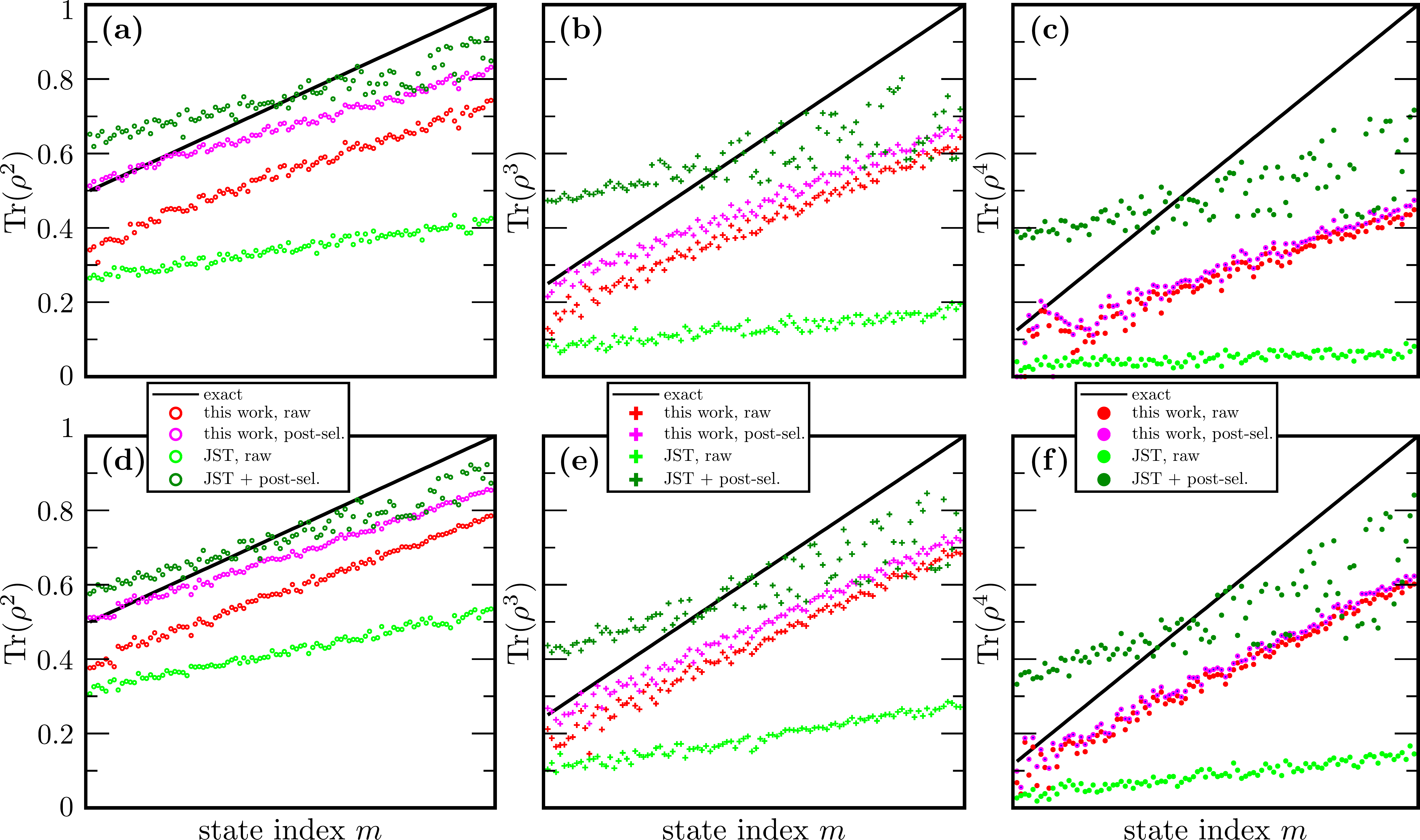}
\caption{Numerical simulations with and without post-selection. Plots (a)-(c) show the effect of post-selection in the presence relaxation, decoherence, readout, and gate errors. Plots (d)-(f) show how the algorithms are effected by gate noise alone. All the noise parameters can be found in Appendix~\ref{sec:para}.}
\label{fig:numerics_postsel}
\end{center}
\end{figure*}

\section{Numerical Simulation}\label{sctnumerics}

We employed IBM's QASM Simulator~\cite{qasm} to implement the algorithm presented in Fig.~\ref{fig:our_alg} as well as the JST algorithm. We considered $k=1$, so that $\rho_A = \Tr_B (\dya{\psi} )$ is a one-qubit state, and $n=2$, $n=3$, and $n=4$ corresponding to $\Tr(\rho_A^2)$, $\Tr(\rho_A^3)$, and $\Tr(\rho_A^4)$. For the state $\ket{\psi}$, we generated 100 states denoted $\ket{\psi_m}$ with $m=1,\ldots,100$. These states were prepared as follows:
\begin{align}\label{eqn18}
\ket{\psi_m} = V_B V_A\text{CNOT}_{AB} R^y_A(\theta_m)\, \ket{0}_A \ket{0}_B\,,
\end{align}
where $\text{CNOT}_{AB}$ is a controlled-NOT with $A$ ($B$) as the control (target) qubit, $V_A$ and $V_B$ are Haar random one qubit unitaries, and $R^y_A(\theta_m)$ is a rotation of subsystem $A$ about the $y$-axis. The parameter $\theta_m$ determines how much entanglement $\ket{\psi_m}$ has and was chosen such that $\text{Tr}(\rho_A^n)$ takes values equally spaced between its smallest and largest possible values. 

Our numerical simulations accounted for relaxation and decoherence due to $T_1$ and $T_2$ processes, as well as gate and readout errors. Our simulations account for noise errors that affect both the algorithm as well as the state preparation (i.e., preparing the various copies of $\ket{\psi}$). The noise parameters are given in Appendix~\ref{sec:para}.  
The simulation results are shown in Fig.~\ref{fig:numerics}. Panels (a), (b), and (c) compare the algorithms in Fig.~\ref{fig:our_alg} and Fig.~\ref{fig:troyer} with the exact curve for $n= $ 2, 3, and 4, respectively. In each case, the algorithm in Fig.~\ref{fig:our_alg} gets closer to the exact curve. The relative errors for the two algorithms are shown in panels (d) and (e).

For the particular noise parameters chosen in Fig.~\ref{fig:numerics}, it appears that the error increases with $n$ more rapidly for the algorithm in Fig.~\ref{fig:troyer}. Naturally, one expects the error to increase with $n$ for both algorithms, since the number of state preparations of $\ket{\psi}$ (each of which has an associated error) increases linearly with $n$. In addition, the depth of the algorithm in Fig.~\ref{fig:troyer} increases linearly in $n$, while the number of measurements in the algorithm in Fig.~\ref{fig:our_alg} increases linearly in $n$. As a result, one expects the algorithm in Fig.~\ref{fig:troyer} to be more sensitive to decoherence (due to the increased depth), while the algorithm in Fig.~\ref{fig:our_alg} should be more sensitive to readout error (due to the increased number of measurements).

Based on the gate counts of each algorithm alone we expect gate errors to effect our algorithm less than the JST algorithm. This is because, for each CNOT gate that one needs to implement in our algorithm, one needs to implement 8 CNOT gates in the JST algorithm~\cite{shende2009cnot}. 
Nevertheless, the effect of gate errors is subtle because the two algorithms have a different number of copies of the state. To address this issue, we performed numerical simulations with only gate errors (all other sources of errors being absent). The results are displayed in Fig.~\ref{fig:numerics_postsel}(d)-(f) and show a similar pattern as the simulations in Fig.~\ref{fig:numerics_postsel}(a)-(c), which involved multiple noise mechanisms.

In Fig.~\ref{fig:numerics_postsel} we explore the potential of post-selection methods for improving the accuracy of the results. The correction due to post-selection roughly appears to be a shift of the data upward by a constant, without affecting the slope. The slope, on the other hand, is what allows one to distinguish low entangled states from high entangled states. This means that post-selection may be particularly useful for results that already have close to the correct slope. As can be seen in Fig.~\ref{fig:numerics_postsel}, the algorithm presented in this paper captures the slope of the data much better, especially for higher R\'enyi indices.

\section{Hardware noise versus statistical noise}

Reliable values for the various $R_n$ are critical for faithfully extracting the entanglement spectrum using Eq.~\eqref{eqn11}. 
Above we analyzed the effect of hardware noise on the accuracy of the $R_n$ values. In addition, statistical noise due to finite statistics limits the precision of the $R_n$ values. Here we analyze the statistical noise due to finite sampling and compare it to the hardware noise levels we observed in the numerical simulations.  

In both our's and JST's algorithm, each run results in a single number  $\{\pm 1\}$, which is then averaged over many runs. This means that the standard deviation of the final outcome will scale as $O(1/\sqrt{M})$, where $M$ is the number of runs. (In Ref.~\cite{johri2017entanglement} it was pointed out that the technique of quantum amplitude estimation can be used to improve this scaling.) The JST algorithm outputs $R_n = \text{Tr}(\rho^n)$ with standard deviation $\sigma = \sqrt{R_n(1-R_n)}/\sqrt{M}$. Our algorithm, on the other hand, outputs $\vert \text{Tr}(\rho^n)\vert^2=R_n^2$ with $\sigma = \sqrt{R_n^2(1-R^2_n)}/\sqrt{M}$. From this, $R_n$ can be determined with precision $\sigma = \sqrt{1-R_n^2}/2\sqrt{M}$, which is valid when $M\gg 1/R_n^2$. 

In our numerical simulations, Figs.~\ref{fig:numerics} and \ref{fig:numerics_postsel}, we used $M=10,000$ runs for each data point. The error bars associated with the standard deviations obtained above are too small to be seen on the plot. This means that for the parameters chosen for this simulation, statistical errors are negligible compared to other sources of noise associated with the hardware. We expect this trend to hold true in the NISQ era.

\section{Conclusions}\label{sctconclusion}

In this work we presented a new quantum algorithm for computing the integer R\'enyi entropies, which can be used to determine the entanglement spectrum. Relative to the algorithm in \cite{johri2017entanglement}, our algorithm doubled the number of qubits required but dramatically reduced the circuit depth. 
In doing so, we traded a large circuit depth (which is proportional to $k*n$ for the algorithm in \cite{johri2017entanglement}) for a more expensive classical post-processing (which takes time proportional to $k*n$ in our algorithm). Hence our algorithm is a hybrid quantum-classical algorithm. As a result, the quantum portion of our algorithm has a depth of only two gates. This makes it ideal for implementations on NISQ computers, whose qubit counts are rapidly growing, but whose qubits remain noisy.

We remark that computing higher order R\'enyi entropies may be used to compute bounds for von Neumann entropic quantities \cite{smith2017quantifying}. We also note that the Two-Copy Test in Fig.~\ref{fig2} for computing the expectation value of an operator $M$ may be of independent interest, since it avoids implementing a (costly) controlled-$M$ gate as in the Hadamard Test and hence is more amenable to blackbox implementation of $M$~\cite{araujo2014quantum,thompson2018quantum}.

\vspace{3mm}
\section{Acknowledgements}\label{sctackknowledge}

All authors acknowledge support of the LDRD program at Los Alamos National Laboratory (LANL). LC was also supported by the U.S. Department of Energy through the J. Robert Oppenheimer fellowship. PJC was also supported by the LANL ASC Beyond Moore's Law project.

\bibliography{spec.bbl}

\appendix

\section{Post-selection Methods}

\subsection{Algorithm of Ref.~\cite{johri2017entanglement}}
\label{sec:johripost}

In this Appendix we generalize the post-selection method described in Ref.~\cite{linke2017measuring} for the algorithm of Ref.~\cite{johri2017entanglement}. 
We show that the complexity of the procedure scales as $O(k\, n\log n)$. 
First, we define the following shorthand notation:
\begin{align}
    \balpha &\equiv (\valpha^{(1)},\valpha^{(2)},\dots, \valpha^{(n)})\; ,\\
    \valpha^{(s)} &\equiv \big( \underbrace{\alpha^{(s)}_1,  \dots, \alpha^{(s)}_k}_{\text{subsystem A}}, \underbrace{\alpha^{(s)}_{k+1}, \dots, \alpha^{(s)}_{2k}}_{\text{subsystem B}} \big)\; ,\\
    \bpsi_{\balpha} &\equiv \psi_{\valpha^{(1)}}\psi_{\valpha^{(2)}} \dots \psi_{\valpha^{(n)}}\;,
\end{align}
where $\alpha_j^{(s)}\in\{0,1\}$.
Then, the initial state of the quantum circuit of Fig.~\ref{fig:troyer} can be expressed as 
\begin{align}
    \ket{0}\sum_{\balpha} \bpsi(\balpha) \ket{\balpha} \; .
\end{align}
At the end of the circuit shown in Fig.~\ref{fig:troyer}, and before the measurement, the state is given by
\begin{align}
    \frac{1}{2} \bigg[ \ket{0} \sum_{\balpha}\bpsi_{\balpha} &\left( \ket{\balpha}  + \ket{\perm(\balpha)} \right) \\
    \nonumber
    &+ \ket{1} \sum_{\balpha} \bpsi_{\balpha} \left( \ket{\balpha} - \ket{\perm(\balpha)}\right) \bigg]\; ,
\end{align}
where $\perm$ is defined in Eq.~\eqref{eqn8}. 
The amplitude associated with a given measurement outcome of the form $\ket{0}\ket{\balpha}$ is in general nonzero. However amplitudes for configurations of the form $\ket{1}\ket{\balpha}$ vanish, for all states $\ket{\psi}$, when the following condition is satisfied:
\begin{align}
    \label{eq:cond1}
    \bpsi_{\balpha}  = \bpsi_{\perm^{-1}(\balpha)}\; .
\end{align}
Let $\balpha_*\equiv \perm^{-1}(\balpha)$. We can then write:
\begin{align}
    \balpha_* \equiv (\valpha_*^{(1)}, \valpha_*^{(2)},\dots,\valpha_*^{(n)})\; .
\end{align}
Using this notation Eq.~\eqref{eq:cond1} can be rewritten as
\begin{align}
    \psi_{\valpha^{(1)}}\dots \psi_{\valpha^{(n)}} = \psi_{\valpha_*^{(1)}}\dots \psi_{\valpha_*^{(n)}}\; .
\end{align}
This holds for an arbitrary state if the following holds
\begin{align}
    \label{eq:cond2}
    \{ \valpha^{(1)} ,\dots, \valpha^{(n)} \} = \{ \valpha_*^{(1)} ,\dots, \valpha_*^{(n)} \}\; ,
\end{align}
where the curly brackets indicate a set. 

The post-selection method works as follows. Measure all qubits. If ancilla qubit is in state 0, accept outcome. If ancilla qubit is in state 1, check condition given in Eq.~\eqref{eq:cond2}. If condition is satisfied, discard the outcome, else accept it.

Condition \eqref{eq:cond2} can be checked efficiently using the following procedure. First, sort the entries in each set using a  comparison-based sorting algorithm. 
A comparison operation can be defined in this case, for example, by interpreting $\valpha = (\alpha_1, \dots, \alpha_{2k})$ as the binary representation of an integer $\in [0, 2^{2k}-1]$. 
In the worst case this requires comparing $2k$ binary variables. 
Comparison-based sorting algorithms, such as \textit{merge sort} 
need $O(n \log n)$ calls to the comparison operation. Thus the complexity of sorting is $O(k\,n\log n)$. 

Once we have sorted both sets in Eq.~\eqref{eq:cond2}, comparing them requires comparing each of the $n$ elements. Comparing each element involves comparing $2k$ binary variables. Thus the complexity of comparing sorted lists is $O(k\,n)$. It follows that the overall complexity is dominated by the sorting algorithm and is given by $O(k\,n\log n)$ which is almost linear in $k*n$.

\subsection{Algorithm based on Two-Copy Test}
\label{sec:ourpost}

In this Appendix we discuss the post-selection procedure for our algorithm. We show that the complexity of the procedure scales as $O(k \, n^2)$.

The state of the algorithm just before the measurement is given by
\begin{equation}
    \ket{\Omega} = \sum_{\balpha,\bbeta} \Omega_{\balpha,\bbeta} \ket{\balpha,\bbeta} \ ,
\end{equation}
where $\balpha$ is a multi-index that corresponds to the first set of $n$ copies of $\ket{\psi}$. Similarly, $\bbeta$ refers to the second one. Amplitudes $\Omega_{\balpha,\bbeta}$ read
\begin{equation} \label{eq:omega}
    \Omega_{\balpha,\bbeta} \propto \sum_{\bgamma }
    \Upsilon_{\bgamma }
    \Upsilon_{P(\bgamma) \oplus \bbeta}
    (-1)^{\balpha \cdot \bgamma} \ .
\end{equation}
Here,
\begin{equation}
    \ket{\Upsilon} = \sum_{\bgamma} \Upsilon_{\bgamma} \ket{\bgamma} = \ket{\psi}^{\otimes n} \ ,
\end{equation}
$\oplus$ denotes summation mod 2 and $P$ is the permutation of qubits of $\ket{\Upsilon}$.  Permutation $P$ is defined in Eq.~\eqref{eqn8}. An example of this permutation is given in Fig.~\ref{fig:our_alg}(a)-(b). Let us now find conditions that $\balpha$ and $\bbeta$ need to satisfy for $\Omega_{\balpha,\bbeta} = 0$ to hold for all $\ket{\psi}$.

Changing the summation index in Eq.~\eqref{eq:omega} from $\bgamma$ to $P^{-1}(\bgamma) \oplus \bbeta$ we obtain:

\begin{equation}
\begin{split}
    \Omega_{\balpha,\bbeta} \propto &
    (-1)^{\balpha \cdot \bbeta}
    \sum_{\bgamma}
    \Upsilon_{P^{-1}(\bgamma)\oplus \bbeta}\\
    & \Upsilon_{\bgamma \oplus P(\bbeta) \oplus \bbeta}
    (-1)^{P^{-1}(\balpha) \cdot \bgamma} \ .
\end{split}
\end{equation}

We will use the fact that $\ket{\Upsilon}$ is symmetric under any permutation $T$ of copies $\ket{\psi}$. That is $\Upsilon_{T(\bgamma)} = \Upsilon_{\bgamma}$. There are $n!$ such permutations, but only $n$ of them will be relevant for our post-selection procedure, as we will show below. Changing the summation index to $T(\bgamma)$ we obtain

\begin{equation} \label{eq:omega2}
\begin{split}
    \Omega_{\balpha,\bbeta} \propto &
    (-1)^{\balpha \cdot \bbeta}
    \sum_{\bgamma}
    \Upsilon_{P^{-1}(T(\bgamma))\oplus \bbeta} \\
    & \Upsilon_{\bgamma \oplus T^{-1}(P(\bbeta) \oplus \bbeta)}
    (-1)^{T(P^{-1}(\balpha)) \cdot \bgamma} \ .
\end{split}
\end{equation}

Comparing Eq.~\eqref{eq:omega2} to Eq.~\eqref{eq:omega} we see that the amplitude $\Omega_{\balpha,\bbeta} = 0$ if the following conditions on $\balpha$, $\bbeta$ and permutation $T$ are met

\begin{align}
    P^{-1} \circ T &= T \circ P \, ,\label{eq:PT} \\
    \balpha  \cdot \bbeta &\equiv 1 \pmod 2\, , \\
    T(P^{-1}(\balpha)) &= \balpha\, , \\
    T(\bbeta) &= \bbeta\, ,  \\
    P(\bbeta) &= \bbeta \, .
\end{align}

Because $P$ is a permutation with a single cycle, there are only $n$ permutations $T$ that satisfy condition $\eqref{eq:PT}$. All those permutations can be generated by fixing the mapping of the first element to one of $n$ possible outcomes. The full action of the permutation is then dictated by Eq.~\eqref{eq:PT}. In time $O(n^2)$, we can obtain all permutations $T$ that satisfy Eq.~\eqref{eq:PT}. For every such permutation, the rest of the conditions above can be checked in time $O(k\, n)$, since $2 n\, k$ is the total number of qubits in $\ket{\Upsilon}$.

This shows that, for a given measurement outcome $\balpha$ and $\bbeta$, checking if this is a forbidden outcome ($\Omega_{\balpha,\bbeta} = 0$) takes $O(k \, n^2)$ time.

\section{Parameters of Numerical Simulation}
\label{sec:para}

We used QASM Simulator \cite{qasm} developed by IBM for generating the plots in Sec.~\ref{sctnumerics} of the main text. 
QASM Simulator is a quantum circuit simulator written in Python that includes a variety of realistic circuit level noise models. We used it as a local backend in the Quantum Information Science Kit (QISKit version 0.5.2) Python SDK. 

For each data point we used 10,000 runs. The probability of readout error was set to 0.02. This means that $2\%$ of the time a 0 outcome is interpreted
as 1 and vice versa. 
The relaxation rate has been set to 0.005 in units of a single qubit gate time. The CNOT gate is assigned a duration 5 times that of a single qubit gate. 
The relaxation rate $r$ specifies the $T_1$ and $T_2$ relaxation error of a system (with $T_2=T_1$). The probability of a relaxation error for a gate of length $t$ is given by $p_{\mathrm{err}} = 1-\exp(-t r) $. If a relaxation error occurs the system is reset to the 0 or 1 states with probability $p_0$ and $p_1 = 1-p_0$ respectively, where we chose $p_1=10^{-7}$.

In order to model gate errors we adopted an error model in which the only nontrivial gates are the 90-degree $X$ rotation, denoted $X90$, and the CNOT gate. All single qubit gates are implemented in terms of noisy $X90$ gates and ideal $Z$-rotations. The $X90$ gate is a $90$-degree rotation around $X$ axis, with the matrix representation
\begin{equation}
X90 = \frac{1}{\sqrt{2}}
\left( \begin{matrix}
1 & -i \\
-i & 1
\end{matrix}\right) \ .
\end{equation}
We have included depolarizing and Pauli error channels for the quantum gates. For the one qubit $X90$ gate the depolarizing probability is set to $0.001$, whereas for the CNOT gate this probability is $0.005$. For the one qubit X90 gate all the three Pauli error probabilities are set to $0.001$, whereas for the CNOT gate all the 15 Pauli error probabilities are $0.005$.

\end{document}